\documentclass[aps,prl,reprint,superscriptaddress,floatfix,amsfonts,amssymb,amsmath,
showpacs,showkeys]{revtex4-1}
\usepackage[english]{babel}

\usepackage{amssymb,multirow}
\usepackage{amsmath}
\usepackage{bm}
\usepackage{graphicx}
\usepackage{float}
\usepackage{hyperref}

\providecommand*{\mnras}{{\it MNRAS}}

\providecommand*{\pre}   {{\it Physical Review E}}
\providecommand*{\apj}   {{\it Astrophys. J.}}
\providecommand*{\apjl}   {{\it Astrophys. J. Lett.}}

\renewcommand{\vec}[1]{\mathbf{#1}}

\newcommand{\dif}[1]{\vec\nabla^2#1}

\newcommand{\ex}{\vec{\hat x}}

\newcommand{\mr}[1]{\mathrm{#1}}

\def\rrr#1\\{\par
\medskip\hbox{\vbox{\parindent=2em\hsize=6.12in
\hangindent=4em\hangafter=1#1}}}

\usepackage{color}

\begin{document}

\title[Reconnection Turbulence]{Influence of tearing instability on magnetohydrodynamic turbulence}

\author{Justin Walker}
\email[E-mail: ]{jwwalker2@wisc.edu}
\affiliation{Department of Physics, University of Wisconsin-Madison, 1150 University Avenue, Madison, WI 53706, USA}
\author{Stanislav Boldyrev}
\affiliation{Department of Physics, University of Wisconsin-Madison, 1150 University Avenue, Madison, WI 53706, USA}
\affiliation{Space Science Institute, Boulder, Colorado 80301, USA}
\author{Nuno F.\ Loureiro}
\affiliation{Plasma Science and Fusion Center, Massachusetts Institute of Technology, Cambridge MA 02139, USA}

\date{\today}

\label{firstpage}

\begin{abstract}
  It has been proposed recently by  Loureiro \& Boldyrev [Phys. Rev. Lett. \textbf{118}, 245101 (2017)] and Mallet \emph{et al.} [Mon. Not. R. Astron.
Soc. \textbf{468}, 4862 (2017)]
  that strongly anisotropic current sheets formed in the inertial range of magnetohydrodynamic turbulence become affected by the tearing instability at scales smaller than a critical scale $\lambda_\mr{c}$, and larger than the dissipation scale of turbulence.
  If true, this process can modify the nature of energy cascade at smaller scales, leading to a new, tearing-mediated regime of magnetohydrodynamic (MHD) turbulence.
  In this work we present a numerical study of strongly anisotropic, two-dimensional turbulent eddies, and we demonstrate that the tearing instability can indeed compete with their nonlinear evolution.
  The results, therefore, provide direct numerical support for the picture that a new regime of MHD turbulence can exist below~$\lambda_\mr{c}$.
\end{abstract}

\maketitle

\section{Introduction}
Plasma turbulence occurring in natural systems, such as the interstellar medium, the solar corona, solar wind, planet magnetospheres, etc., typically spans a very broad range of scales and affects important phenomena like plasma heating, particle acceleration and scattering~\cite[e.g.,][]{biskamp2003,chen2016}.
At scales larger than the ion kinetic scales, the plasma dynamics can be modeled in the framework of magnetohydrodynamics \cite[e.g.,][]{biskamp2003,chen2016,tobias2013,davidson2017,chen2017}.
Magnetohydrodynamic turbulence can be viewed as nonlinear Alfv\'enic modes or eddies propagating along the local background magnetic field.
Such eddies are expected to be anisotropic with respect to the background field \cite[e.g.,][]{goldreich_toward_1995}.
Moreover, they assume the shapes of ribbons or current sheets at progressively smaller scales \cite[e.g.,][]{boldyrev2005,boldyrev_spectrum_2006,chen_3D_2012,chandran_15,mallet_measures_2016}.
This picture is consistent with (and may provide an explanation for) the numerically  observed morphology of small-scale current structures in magnetohydrodynamic (MHD) turbulence \cite[][]{matthaeus_turbulent_1986,biskamp2003,servidio_magnetic_2009,servidio_magnetic_2011,wan2013,zhdankin_etal2013,zhdankin_etal2014}. 

Given a very large Reynolds number, the ribbon-like eddies in the inertial interval of MHD turbulence may become affected by the tearing instability.
This question was first addressed in~\cite{carbone1990} in the framework of the Iroshnikov-Kraichnan (IK) model of MHD turbulence~\cite[][]{iroshnikov_turbulence_1963,kraichnan_inertial_1965}.
The tearing mode considered in~\cite{carbone1990} was essentially isotropic, which fit an assumption of the IK model that MHD turbulence consists of isotropic (characterized by a single size) weakly interacting Alfv\'en waves at each scale
\footnote{The IK model has since been demonstrated to be incorrect, since weak MHD turbulence has the energy spectrum $-2$, not $-3/2$ proposed in the IK model \cite[e.g.,][]{ng1996,goldreich1997,galtier2000}.}.
Moreover, the tearing mode considered in~\cite{carbone1990} required the presence of a significant velocity shear tuned to the magnitude and scale of the magnetic field \cite[][]{hoffman1975,dobrowolny1983,einaudi1986}.
The growth rate of this mode depended on the velocity shear, and it reduced to the standard Furth-Killeen-Rosenbluth (FKR) result \cite{FKR} as the velocity shear decreased \cite{einaudi1986}.

In \cite[][]{loureiro2017} and \cite{mallet2017} it was proposed that MHD turbulence should rather be modified at small scales by highly anisotropic tearing modes, which are beyond the FKR regime.  
It was conjectured that a new, tearing-mediated energy cascade  should exist in the range of scales intermediate between the Alfv\'enic inertial interval and the dissipation scale of MHD turbulence.
The transition scale to the tearing-mediated regime depends on the model shape assumed for the turbulent eddies.
If the sheared magnetic structures have a ``tanh-like'' profile \cite[][]{harris_1962}, the scale is given by $\lambda_\mr{c}\sim LS_L^{-4/7}$, where $L$~is the outer scale of turbulence and $S_L$~the corresponding Lundquist number.
For a ``sine-like'' profile that is arguably more appropriate for turbulent systems and that we study in this work, the transition scale is estimated slightly differently, $\lambda_\mr{c}\sim LS_L^{-6/11}$~\cite[][]{Boldyrev_2017} \footnote{It can be demonstrated that the tearing mode pertinent to our study grows faster than the mode of the tanh-like profile and the mode considered in \citet{carbone1990}}.

It was estimated that such a regime becomes relevant if the magnetic Reynolds number%
\footnote{We assume that at the outer scale the magnetic fluctuations are comparable to the velocity fluctuations and to the background magnetic field.
  We also assume that the fluid viscosity and magnetic diffusivity are on the same order, so the Lundquist, Reynolds, and magnetic Reynolds numbers are on the same order as well.} of turbulence becomes very large $\mr{Rm} \gtrsim 10^6$ \cite[e.g.,][]{Boldyrev_2017}.
Due to this severe computational constraint, direct numerical evidence in support of the tearing-mediated turbulence regime does not exist.%
\footnote{It has been observed previously that the tearing instability initiated in a thin laminar current layer eventually leads to a broad range of nonlinearly interacting and seemingly turbulent fluctuations \cite[][]{huang_turbulent_2016,hu2018}.
In those works, however, the reconnection layer did not possess Alfv\'enic turbulence.}

In this work, we propose a method for studying anisotropic MHD turbulence in the tearing-mediated interval with a two-dimensional setup that models the transverse dynamics of a current sheet.
Our method is somewhat analogous to the reduced-MHD approach (RMHD) in simulations of MHD turbulence \cite[e.g.,][]{kadomtsev_p74,strauss_nonlinear_1976,biskamp2003,tobias2013}.
The RMHD equations apply when the simulation domain (a rectangular box) is permeated by a strong background magnetic field $B_0$, say in the $z$-direction.
Assume that the rms values of magnetic and velocity fluctuations are normalized to unity, $v_\mr{rms}\sim b_\mr{rms}\sim 1$.
For the turbulence to be critically balanced at the largest scale, one needs to elongate the box in the $z$-direction proportionally to the value of $B_0$.
In the case $B_0\gg b_\mr{rms}$, the fluctuations of the $z$-components of the magnetic and velocity fields can then be neglected, and the MHD system is approximated by the reduced-MHD equations (see also \cite[][]{oughton2017,zhdankin2017}).

The novelty of our approach is that instead of studying turbulence driven at large scales, we study the evolution of a {\em particular highly anisotropic eddy} that is expected to exist at scales much smaller than the outer scale of the turbulence.
For that we stretch the box in the $x$-direction as compared to the $y$-direction, $L_x \gg L_y$.
For the eddy to be critically balanced, we need the following conditions at the box scale: $L_z/B_0\sim L_x/b_x\sim L_y/b_y$, where $b_x\sim v_x$ and $b_y\sim v_y$ are typical fields in the $x$ and $y$ directions.
The box-sized eddies in such turbulence are effectively very anisotropic current sheets.
It is important to note that such eddies  cannot be in a steady state; they are destroyed by nonlinear interaction on their Alfv\'enic time scale~$\tau_A\sim L_x/b_x$.
During their life time, however, they tend to develop small-scale turbulence inside them that, for a sufficiently large Reynolds number, should resemble regular, although very anisotropic, MHD turbulence.

If we increase the resistivity, however, the large-scale magnetic fluctuations will become subject to tearing instability \cite[e.g.,][]{FKR,coppi_resistive_1976,loureiro_instability_2007}.
The analysis of \cite[][]{Boldyrev_2017} shows that the fastest-growing tearing mode in such an eddy has the growth rate $\gamma_t\sim (b_x/L_y)S^{-3/7}$, where the {\em local, eddy-scale} Lundquist number is defined as $S=b_xL_y/\eta$ and the magnetic field is measured in Alfv\'enic units.%
\footnote{Here we use the fact that the reconnecting magnetic field in our numerical setup has a sine profile.
For a $\tanh$-profile, the exponent $3/7$ should be replaced by $1/2$, which does not qualitatively change the results.}
In order for the tearing rate to become comparable to the eddy turnover rate $\gamma\sim 1/\tau_A \sim b_x/L_x$, we need to require $S= S_{\mr c}\sim (L_x/L_y)^{7/3}$.
Therefore, if we need to perform computations with a large Lundquist number $S$, we have to choose a very anisotropic box. 

On the other hand, in order to reliably measure the scaling properties of the turbulence, the Reynolds number should be large.
The {\em local} Reynolds number measuring the strength of the nonlinear interaction is defined as $\mr{Re}=b_yL_y/\eta$.
It is smaller than the Lundquist number.
For critically balanced fluctuations $b_y\sim b_x(L_y/L_x)$, the Reynolds number corresponding to $S_\mr{c}$ would thus be $\mr{Re}_\mr{c} \sim (b_y/b_x)S_\mr{c}\sim (L_x/L_y)^{4/3}$.
The Alfv\'enic evolution time $\tau_A$ of such an eddy increases with the box elongation.
If we assume that in order to resolve the inertial interval we need at least $\mr{Re}\sim 2000$, and $N_y=512$ grid points in the shortest, $L_y$ direction (see, e.g., \cite[][]{perez_etal2012}), we encounter prohibitively strong limitations for the numerical simulations, in both the number of grid points and the running time. 

In an attempt to overcome these limitations, we use a simplified, two-dimensional setup.
Although two-dimensional MHD is different from its three-dimensional counterpart, there are certain similarities between strong turbulence in the two cases.
As observed numerically \cite[e.g.,][]{politano1998,biskamp_schwarz2001,ng2007,wan2013}, two-dimensional turbulence tends to form sheet-like magnetic structures at small scales, and its energy spectrum is close to~$-3/2$, similar to the three-dimensional case.
The eddy turnover rate should therefore scale in the same way as in three-dimensional turbulence.
We believe that this should suffice for our study of the interplay of tearing and Alfv\'enic dynamics, at least on  a qualitative level.


\section{Numerical method}
We solve the incompressible MHD equations in a two-dimensional anisotropic periodic box with the pseudospectral code \textsc{snoopy}~\cite{lesur2007}.
The equations are
\begin{align}
  \partial_t \vec{v} &= -(\vec{v} \bm{\cdot} \vec{\nabla})\vec{v} -\vec{\nabla} P + \vec{B}\bm{\cdot}\vec{\nabla}{\vec{B}} + \nu \dif{\vec{v}} + \vec f\label{eq:mom},\\
  \partial_t \vec{B} &= \vec{\nabla} \bm{\times} (\vec{v} \bm{\times} \vec{B}) + \eta \dif{\vec{B}} \label{eq:induct},
\end{align}
where $\vec v(x,y,t)$ is the velocity field, $\vec B(x,y,t)=b_0\sin(2\pi y+\phi){\ex}+{\vec b}(x,y,t)$ is the magnetic field, $P$ is the pressure, and $\vec f(x,y,t)$ is the external force.
The magnetic field is measured in Alfv\'{e}nic units, $v_A=B/\sqrt{4\pi\varrho}$.
The large-scale magnetic field $b_0\sin(2\pi y+\phi){\ex}$ is not an exact solution of the resistive MHD equations, therefore the $k_yL_y/(2\pi)=\pm 1$ components of the magnetic field can change in time.
We, however, update these particular components at each time step to ensure that the amplitude $b_0$ does not change.
The dimensionless pressure $P$ ensures the incompressibility of the flow.
For simplicity, we choose $\mr{Pm}= \nu/\eta = 1$.
We normalize the variables in such a way that $L_y=1$, and $b_0\sim 1$.
The time is measured in units of~$L_y/b_0$.

Currently, the exact dynamics of current sheet formation in MHD turbulence is not well understood.%
\footnote{A mechanism of selective decay, related to the cross-helicity conservation may however be at play \cite[e.g.,][]{tobias2013}.} The fluctuations inside our anisotropic eddy therefore are excited from zero level by an eddy-scale driving force.
The amplitude of the anisotropic, solenoidal random force ${\vec f}(x,y,t)$ is chosen to ensure $v_x\sim v_\mr{rms} \lesssim b_0$; the box anisotropy requires $f_y\sim f_x (L_y/L_x)$.
The force is applied in Fourier space; we force the modes $k_xL_x/(2\pi)=\pm 1,\pm 2$, $k_yL_y/(2\pi)=\pm 1,\pm 2$, with amplitudes drawn from a normal distribution and refreshed independently on average every $\tau_f \sim 1$ (a time short compared to the Alfv\'enic time of the eddy).

We simulate a strongly anisotropic eddy with dimensions $L_x \times L_y = 64 \times 1$.
It is interesting to point out that in isotropically driven MHD turbulence, such structures are expected to exists at scales $\sim 10^{7}$ times smaller than the outer scale of turbulence.
We choose the numerical resolution of $N_x \times N_y = 32768\times 512$ grid points.
As discussed above, in order for the tearing-instability rate to match the eddy-turnover rate, the {\em local} Lundquist number, defined as $S=b_0 L_y/\eta$, should satisfy $S \lesssim (L_x/L_y)^{7/3}\sim 14000$, while for $S\gg 14000$, the turbulence is expected to resemble the standard MHD turbulence~\cite[][]{Boldyrev_2017}. 


\section{Results}
We performed three simulations which differ only in the value of the Lundquist number: $S=64000$, $16000$, and~$4000$.
It is important to note that if tearing were irrelevant the Lundquist number would not affect the time it takes to disrupt the eddy.
\begin{figure}
  \includegraphics{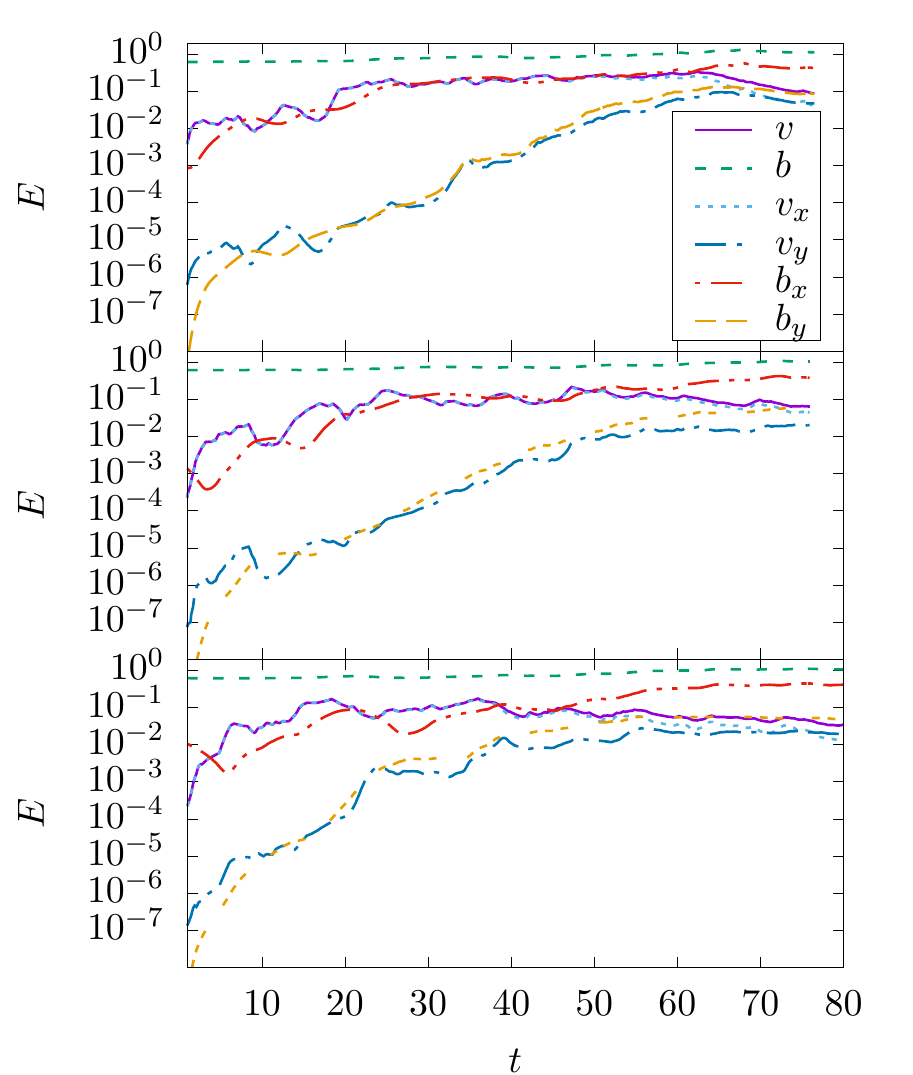}
  \caption{
  Time history of energy components.
  Top panel: $S=64000$, middle: $S=16000$, bottom: $S=4000$.
  The fluctuating $v_y$ and $b_y$ fields are initially generated by the driving force at the level corresponding to $1/64$ of their $x$-components.
  They grow due to nonlinear energy redistribution and/or tearing instability until they reach the magnitude of the $x$-components, at which point the anisotropic eddy is destroyed.}
\label{fig:time-multi}
\end{figure}

\begin{figure}
  \includegraphics{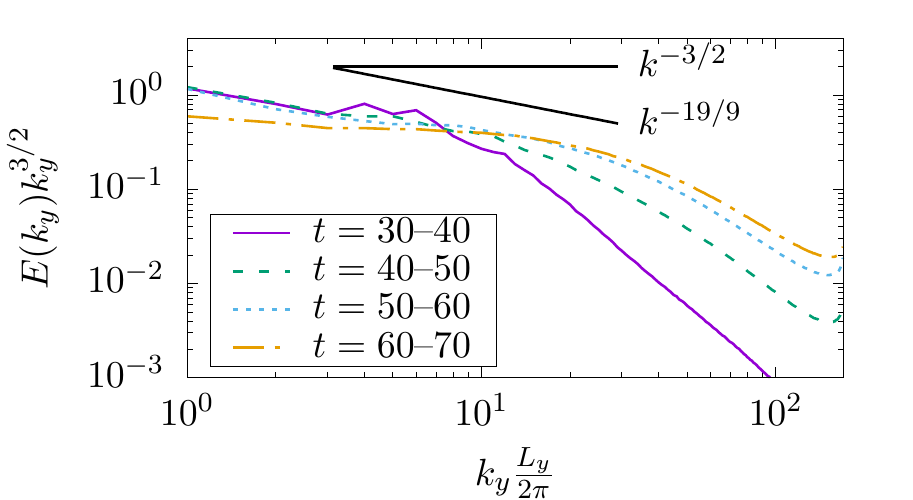}
  \caption{Compensated energy spectrum for the setup with $S=64000$, shown at several instances of the eddy evolution.
    At the latest time interval the eddy has been destroyed by the nonlinear interaction.
    The spectrum at this stage is close to the spectrum of steady-state Alfv\'enic MHD turbulence.}
   \label{fig:spect-64000-late}
\end{figure}

\begin{figure*}
  \includegraphics{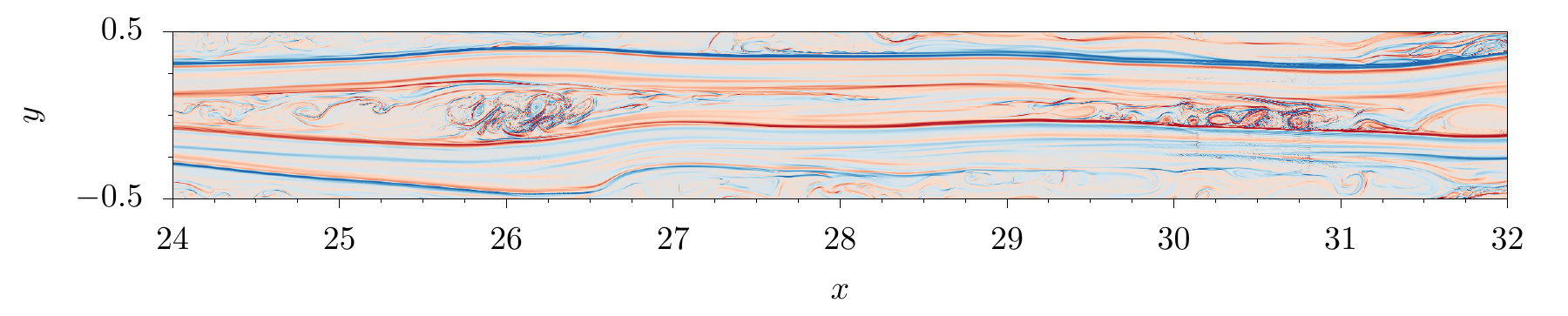}
  \includegraphics{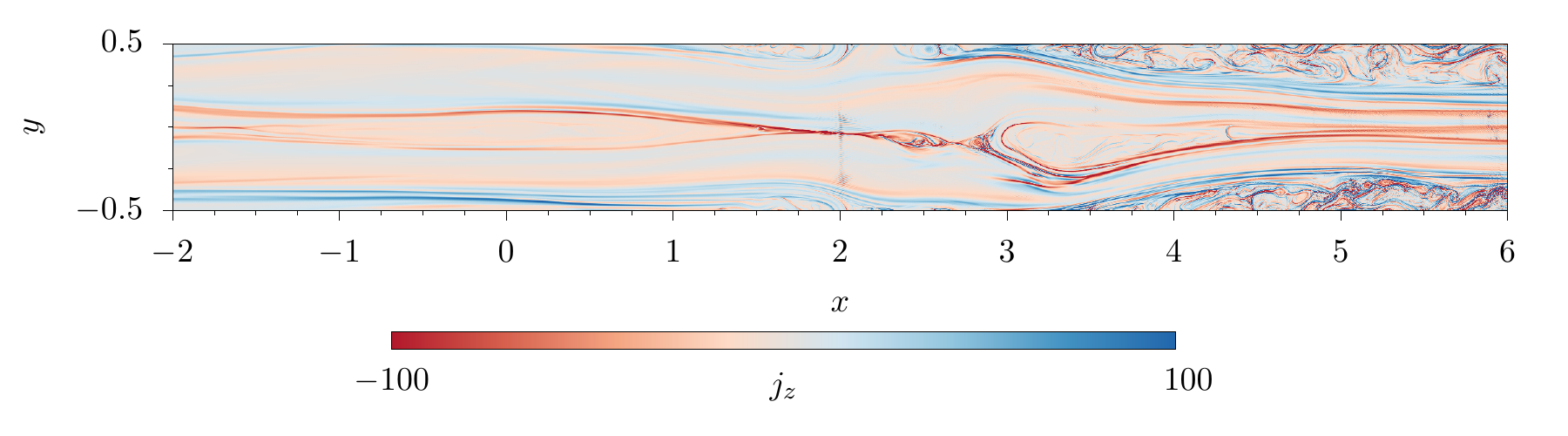}
  \caption{Typical contours of the current $j_z$ for $S=64000$.
    The top panel shows a section of the domain at $t=41$, the bottom one a different section at $t=45$.
    The plasmoid-like structures are not common in the flow; when present they do not fully develop, due to Alfv\'enic shearing flows.}
  \label{fig:64000_snap}
\end{figure*}

Consider, first, the case of the largest Lundquist number $S=64000$.
As seen in Fig.~\ref{fig:time-multi}~(first panel), the anisotropic eddy is gradually destroyed by growing fluctuations of the $b_y$ and $v_y$ fields.
The growth is slow, with a time scale comparable to the Alfv\'enic time scale, $\tau_A\sim 65$.
This time is shorter than the tearing time estimated as $\tau_t\sim (L_y/b_0)S^{3/7}\sim 115$.
It is, therefore, expected that the tearing effects are not important, and indeed the spectrum of the turbulence developing inside the eddy during the eddy evolution is more consistent with that observed in Alfv\'enic turbulence ($-3/2$) in~\cite[e.g.,][]{maron_g01,haugen_04,muller_g05,mininni_p07,chen_11,mason_cb06,mason_cb08,perez_b10_2,perez_etal2012,chandran_15,perez_etal2014} than with the prediction for the tearing-dominated turbulence ($-19/9$), as is shown in Fig.~\ref{fig:spect-64000-late}.
Typical current structures in this case are shown in Fig.~\ref{fig:64000_snap}.
Plasmoidlike structures are not very common.
Even when they appear, they do not have a chance to survive or grow to large scales.
This is consistent with the expectation that the shearing flows associated with Alfv\'enic fluctuations tend to impede the tearing activity.

\begin{figure}
  \includegraphics{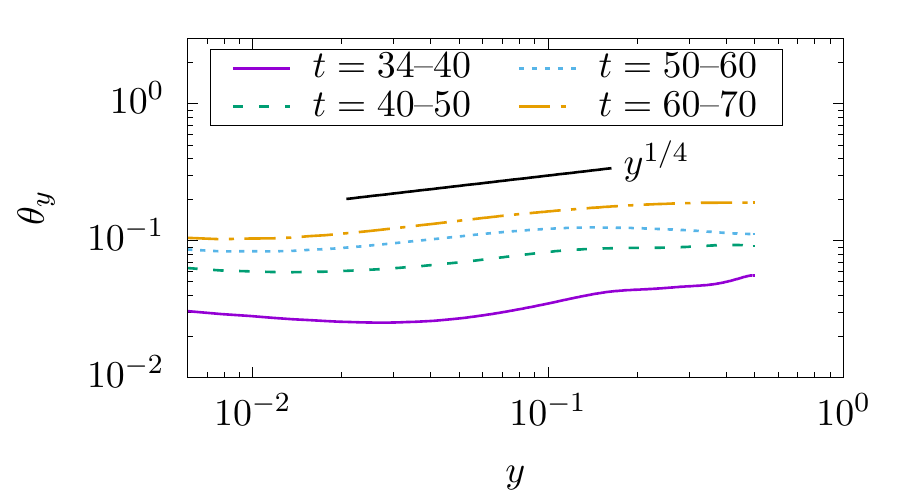}
  \caption{Alignment angle as a function of the short coordinate $y$ for the setup $S=64000$ averaged over different periods of the eddy's evolution.}
  \label{fig:alignment}
\end{figure}

The scaling of the alignment angle between the magnetic and velocity fluctuations, defined as $\theta_\lambda=\sin^{-1}\left(\langle \delta {\bf v}_\lambda\times \delta {\bf b}_\lambda\rangle/\langle |\delta v_\lambda||\delta b_\lambda| \rangle\right)$ (see, e.g., \cite[][]{mason_cb08} for more details), is also broadly consistent with MHD turbulence, even though its overall magnitude changes during the eddy evolution, as shown in Fig.~\ref{fig:alignment}.

The case of $S=4000$ is shown in the bottom panel of Fig.~\ref{fig:time-multi}.
The Lundquist number is small enough so that the tearing time, $\tau_t\sim 35$, is shorter than the Alfv\'enic time.
Therefore, we would expect the eddy to be disrupted faster than in the top panel ($S=64000$), due to the action of the tearing instability.
This observation is consistent with the conjecture (and may serve as proof of the principle) put forward in \cite{loureiro2017,mallet2017,Boldyrev_2017} that the tearing instability can compete with the Alfv\'enic evolution of very anisotropic eddies%
\footnote{The energy spectrum in rather narrow in this case of low~$S$; it does not exhibit a power-law scaling and is not shown here.}.

\begin{figure}
  \includegraphics{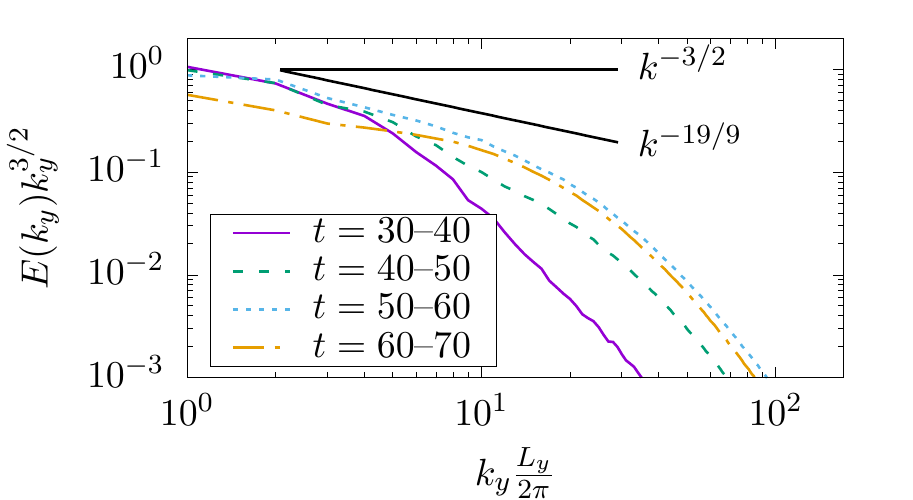}
  \caption{Compensated energy spectrum for setup ${S}=16000$ for several intermediate moments during the eddy evolution.
    The spectrum broadens in $k$-space and seems to approach the slope of $-19/9$, consistent with the prediction for the tearing-mediated turbulence, before the eddy is destroyed at late times.}
   \label{fig:spect-16000}
\end{figure}

\begin{figure*}
  \includegraphics{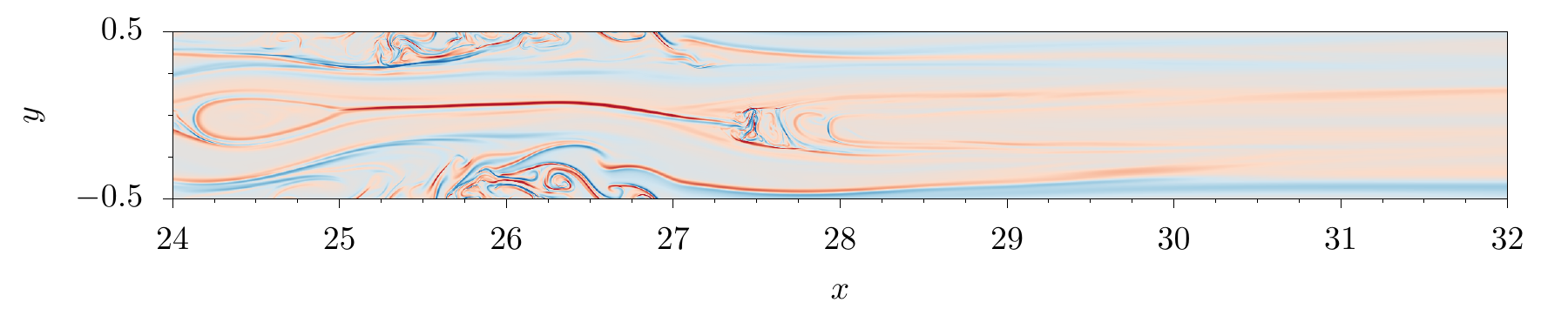}
  \includegraphics{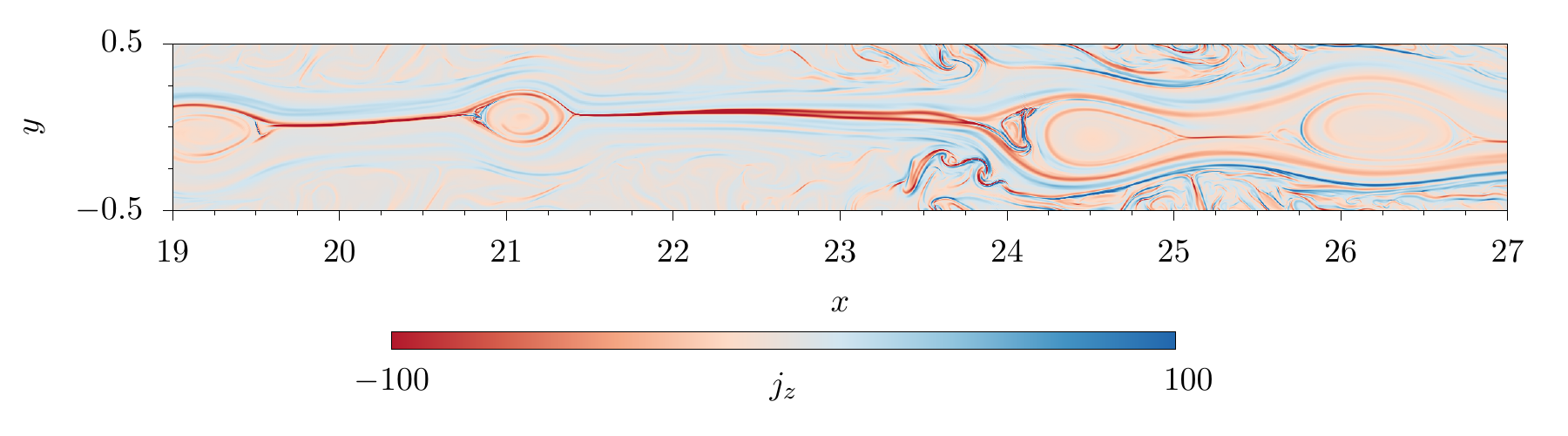}
  \caption{Typical contours of the current $j_z$ for $S=16000$.
    The top panel shows a section of the domain at $t=40$, the bottom one a different section at $t=45$.
    The tendency of turbulence to create plasmoid-like structures is more pronounced as compared to the case depicted in Fig~\ref{fig:64000_snap}.}
  \label{fig:16000_snap}
\end{figure*}

Finally, in the middle panel of Fig.~\ref{fig:time-multi} we show the case $S=16000$ where the Alfv\'enic and tearing times are comparable.
The energy evolution is similar to that in the case of $S=64000$, although the saturation of the growing $y$-components seems to start at a slightly earlier time, in accordance with the increasing importance of the tearing process.
This case is especially important for our consideration.
The energy spectrum of the fluctuations is shown in Fig.~\ref{fig:spect-16000} for several different instances during the eddy evolution.
We observe that as the turbulence is developing inside the eddy, its spectrum broadens in $k$-space and approaches a slope close to $-19/9$, consistent with the prediction for the tearing-mediated turbulence \cite[][]{Boldyrev_2017}.
In this case, the tearing instability has a better chance to compete with the Alfv\'enic fluctuations.
The more pronounced plasmoid-like current structures observed in this case -- see Fig.~\ref{fig:16000_snap} -- strengthen this interpretation.
At the very late stages of the eddy evolution, when the anisotropic eddy is destroyed, the spectrum of the resulting steady-state fluctuations seems to be approaching the shallower $-3/2$ spectrum of regular MHD turbulence.

\begin{figure}
  \includegraphics{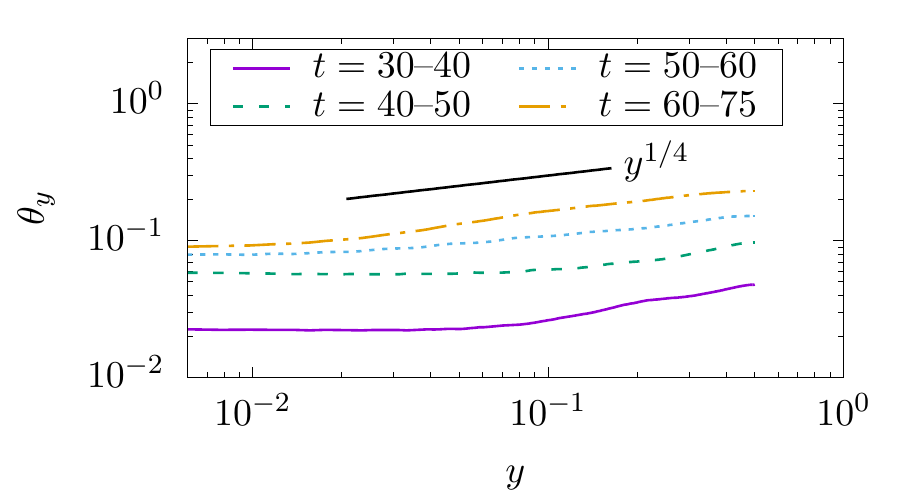}
  \caption{Alignment angle for setup ${S}=16000$ averaged over different periods of the eddy's evolution.}
   \label{fig:align-16000}
\end{figure}

The alignment angle measured for the case of $S=16000$ however shows a difference with the predictions of \cite[][]{Boldyrev_2017}.
Fig.~\ref{fig:align-16000} shows that the alignment angle does not increase at small scales, as predicted in \cite[][]{Boldyrev_2017}.
The reason for that is presently not clear.
It may be related to the principal differences between the 2D and 3D cases, to the limited Reynolds number, or it may indicate that the assumption of Alfv\'enization of tearing-mediated turbulence made in \cite[][]{Boldyrev_2017} is incorrect.

\section{Conclusions}
It has been proposed in~\cite[][]{loureiro2017,mallet2017,Boldyrev_2017} that tearing instability can play a significant role in the inertial interval of magnetic turbulence at small scales.
Very recently, detailed analytical and observational studies of this phenomenon have been  conducted~\cite[][]{mallet2017a,loureiro2017a,comisso2018,vech2018}.
In this work, we have presented a numerical study of an interplay between Alfv\'enic and tearing instabilities in MHD turbulence.
Our results indicate that the tearing instability can indeed modify the dynamics of highly anisotropic turbulent eddies.
In agreement with the analytic predictions, this process can lead to a new regime of MHD turbulence at scales larger than the dissipation scale. 

\paragraph{Acknowledgments.}
J.W. and S.B. were partly supported by the NSF Grant No. NSF PHY-1707272 and NASA Grant No. 80NSSC18K0646.
S.B. was also supported by the Vilas Associates Award from the University of Wisconsin--Madison.
N.F.L. was supported by the NSF-DOE Partnership in Basic Plasma Science and Engineering, Award No. DE-SC0016215 and by the NSF CAREER Award No. 01654168.

\bibliography{master_bl}

\end{document}